\documentclass[twocolumn,traditabstract]{aa}
\usepackage[varg]{txfonts}

\usepackage{graphicx}
\usepackage{natbib}
\usepackage{color}
\usepackage{subfig}
\usepackage[switch]{lineno} 
\usepackage{xspace}
\usepackage{microtype}
\usepackage{amssymb}
\def\link_col{blue}
\usepackage{xspace}
\usepackage{url}
\usepackage{wasysym}
\usepackage{caption}

\def\gray{$\gamma$-ray\xspace}

\begin{document}

\title{On the energy distribution of relativistic electrons in the young supernova remnant G1.9+0.3}
\titlerunning{electron of SNR G1.9+0.3}
\author{Rui-zhi Yang\inst{1}
\and Xiao-na Sun\inst{1}
\and Felix Aharonian\inst{1,2,3,4}}
\institute{Max-Planck-Institut f{\"u}r Kernphysik, P.O. Box 103980, 69029 Heidelberg, Germany.
\and Dublin Institute for Advanced Studies, 31 Fitzwilliam Place, Dublin 2, Ireland.
\and  Gran Sasso Science Institute, 7 viale Francesco Crispi, 67100 L'Aquila (AQ), Italy
\and MEPHI, Kashirskoe shosse 31, 115409 Moscow, Russia
}
\date{Received:  / Accepted: } 

\abstract {The broad-band X-ray observations of  
the  youngest known galactic supernova remnant, G1.9+0.3, 
provide unique information about the particle acceleration at the early 
stages of evolution of  supernova remnants.  Based on  the publicly available X-ray 
data obtained with the Chandra and NuSTAR  satellites over two decades in energy,
we derived the energy distribution of relativistic electrons under the assumption
that  detected X-rays are of entirely synchrotron origin.  The acceleration of electrons
was found to be an order of magnitude slower than the 
maximum rate provided by the shock acceleration in  the nominal  Bohm diffusion regime.  
We discuss the implications of this result in the context of  contribution of SNRs to the Galactic Cosmic Rays at PeV energies. }

\keywords{X-rays: individuals: SNR G1.9+0.3; Acceleration of particles; cosmic rays}

\maketitle

\section{Introduction}
Supernova Remnants (SNRs) are believed to be the sites where the bulk of Galactic Cosmic Rays (CRs) are accelerated
up to PeV energies ($1~ \rm PeV=10^{15}~\rm eV$) \citep[see, e.g, ][]{Hillas2013,blasi13}.  In recent years, significant progress has been achieved in a few directions of exploring the CR  acceleration 
in  SNRs, in particular using the \gray observations in the MeV/GeV and TeV energy bands   \citep[see, e.g., ][]{Aharonian2013}. In particular, the detection of the  so-called 
$\pi^0$-decay bump in the spectra of several mid-age SNRs, is considered as a substantial evidence of acceleration of protons and nuclei in SNRs. 
Moreover, the detection of more than ten young (a few thousand years old or younger) 
SNRs in TeV $\gamma$-rays highlights these objects as efficient particle accelerators,  although the very origin of 
$\gamma$-rays (leptonic or hadronic?) is not yet firmly established.  More disappointingly, so far all TeV emitting SNRs do not show energy spectra which would continue as a hard power-law beyond 10 TeV.  For a  hadronic origin of  detected $\gamma$-rays,  the "early" cutoffs in the energy spectra of 
$\gamma$-rays  around or below 10 TeV imply
a lack of protons inside the shells of SNRs with energies significantly larger than 100 TeV, and,  consequently,  SNRs do not operate as PeVatrons.  However, there are two possibilities 
would allow us to  avoid such a dramatic, for the current paradigm of Galactic CRs, conclusion:

\vspace{1mm}

\noindent
(i) The detected  TeV gamma-rays are of  
leptonic (Inverse Compton) origin.  Of course, alongside with the relativistic electrons,  
protons and nuclei can (should) be  accelerated
as well, but we  do not see the related $\gamma$-radiation because of their ineffective 
interactions  caused by the low density of ambient gas;  

\vspace{2mm}

\noindent
(ii) SNRs do accelerate protons to PeV energies, however it occurs at early stages of evolution of SNRs when the shock speeds exceed  10,000 km/s; we do not see the  corresponding radiation well above 10 TeV
because the PeV  protons already have left the remnant.

\vspace{2mm} 

Both these scenarios significantly limit the potential of gamma-ray observations for the search for  CR  PeVatrons.  Fortunately,  there is another radiation component which 
contains an independent and complementary information about  these  extreme accelerators.  It is related to the synchrotron radiation of accelerated electrons,
namely to the shape of the energy spectrum of radiation in the cutoff region 
which can serve as a distinct 
signature of the acceleration mechanism and its efficiency.  In the shock acceleration 
scheme,  the maximum energy  of accelerated particles, $E_0 \propto B \ v_{\rm sh}^2$.  Therefore,  the epoch  of first several hundred years of  
evolution of a SNR, when the  shock speed $v_{\rm sh}$  
exceeds 10,000 km/s and the  magnetic field is large,  $B \gg 10 \ \mu$G, 
could be  an adequate stage for operation of  a SNR as a PeVatron, 
provided, of course, that   the shock acceleration proceeds close to the Bohm diffusion limit \citep[see, e.g., ][]{Misha}. 
Remarkably, in this regime,  the cutoff energy in the synchrotron radiation of the shock-accelerated 
electrons is determined  by a single parameter,  $v_{\rm sh}^2$ \citep{AhAt99,zirakashvili07}. Therefore, for the known shock speed, the position of the cutoff contains an unambiguous information about the acceleration efficiency.  For 
$v_{\rm sh} \simeq 10,000$~km/s,
the synchrotron cutoff in the spectral energy distribution (SED) 
is expected around 10~keV.  Thus,  the  study of  synchrotron radiation in  
the hard X-ray band can shed light on the acceleration efficiency of electrons, and, consequently, 
provide  an answer whether these objects can operate as CR PeVatrons, given that in the shock acceleration scheme the acceleration of electrons and protons is expected to be identical.   In this regard,  
G1.9+0.3, the youngest  known SNR  in our Galaxy \citep{reynolds08, green08},  is a perfect object  to explore this unique tool.   

The X-ray observations 
with the Chandra and NuSTAR satellites \citep{reynolds09, zoglauer15} cover a rather broad energy interval which is crucial for the study of the spectral shape of  synchrotron radiation in the cutoff region.  Such a study has been conducted  by the team of the NuSTAR 
collaboration \citep{zoglauer15}.  However, 
some  conclusions and statements of  that paper seem to us rather confusing  and, to a certain extent,  misleading. 

In this paper we  present  the results of our own analysis of  the NuSTAR and Chandra data
with an emphasis on the study of the SED of X-radiation 
over two decades,  from 0.3 keV to 30 keV.  Using the synchrotron spectrum and the Markov Chain Monte Carlo (MCMC) technique,   we derive the energy distribution of electrons responsible for X-rays, and  discuss the astrophysical implications of the obtained results.

\section{X-ray observations}\label{sec:data}
The recent hard X-ray observations of G1.9+0.3 by the  NuSTAR satellite are unique for understanding of the acceleration and radiation processes of ultrarelativistic electrons 
in SNRs at the early stages of their evolution.  The detailed study of the NuSTAR data, combined with the Chandra observations at lower energies, have been comprehensively analysed by \citet{zoglauer15}. In particular, it was found that the source can be resolved into two bright limbs with similar spectral features.  The combined Chandra and NuSTAR data sets have been claimed to be best described by the so-called 
{\it srcut} model \citep{Reynolds2008} or by the power-law function with an exponential cutoff. The characteristic cutoff energies in these two fits have been found around 3 keV and 15 keV, respectively  \citep{zoglauer15}. 

To further investigate the features of the X-ray spectrum in the 
cutoff region we performed an independent study based on the publicly available Chandra and NuSTAR X-ray data. For NuSTAR, we used  the set of three observations with ID 40001015003, 40001015005, 40001015007, including both the
focal plane  A (FPMA) and B (FPMB) modules.  The data have been 
analysed using the HEASoft version 6.16, which includes NuSTARDAS, the NuSTAR Data Analysis Software package  (the version 1.7.1 with the  NuSTAR CALDB version 20150123). For the Chandra data, we used the ACIS observations with ID 12691, 12692 and 12694.   The Chandra data reduction was performed by using the version 4.7 of the CIAO (Chandra Interactive Analysis of Observations) package.  

\begin{figure*}
\centering
\includegraphics[width=0.4\linewidth]{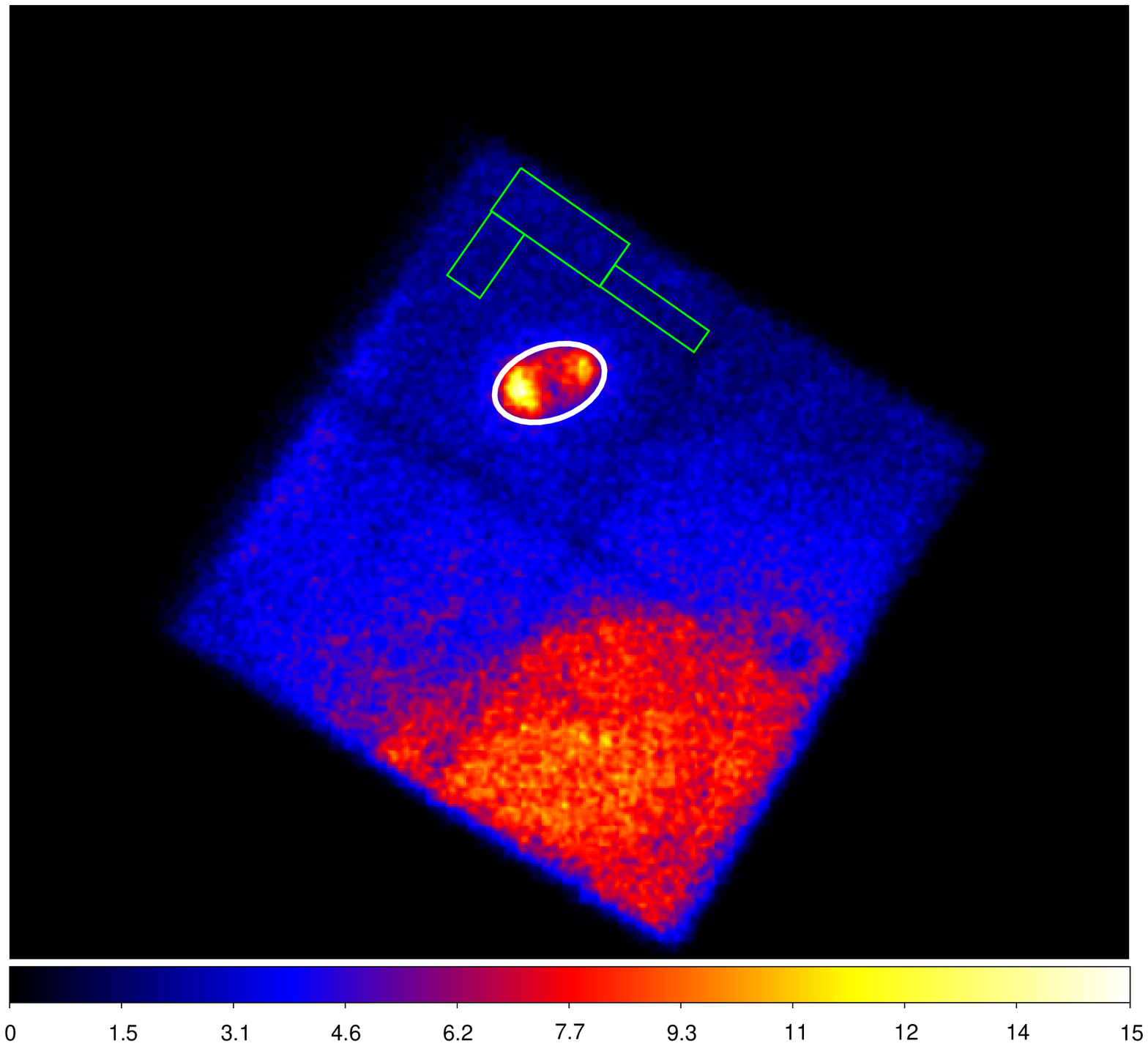}\includegraphics[width=0.4\linewidth]{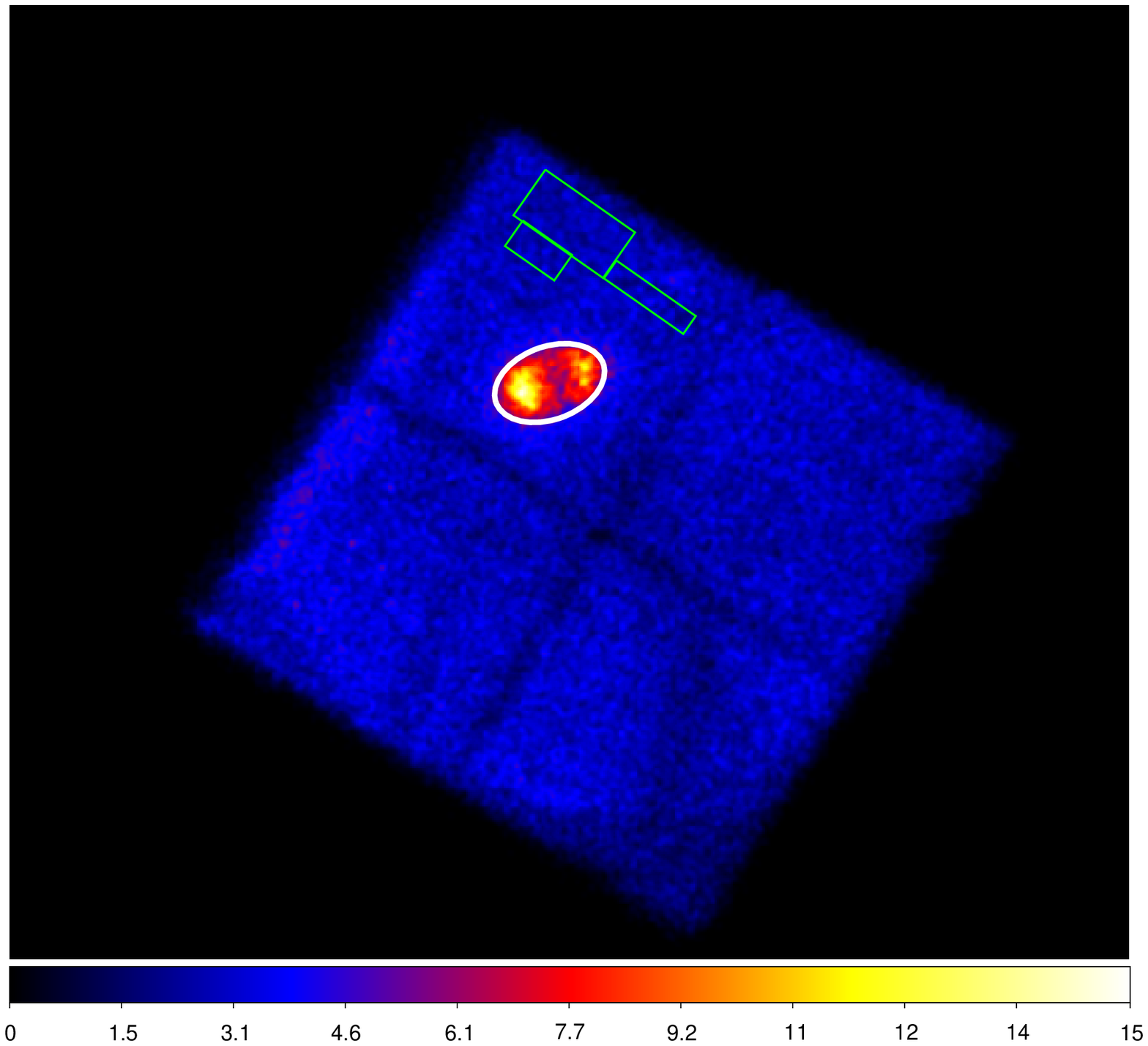}
\caption{Images from the observation 40001015007 for the FPMA (left) and FPMB (right) modules. 
The source and background regions are indicated by the white and green contours, respectively.}
\label{fig:map}
\end{figure*}

In Fig.\ref{fig:map}  we show  the X-ray sky map above 3~keV based on the NuSTAR 
40001015007  data set. In order to  gain from the
maximum possible statistics, for the spectral analysis we have 
chosen the entire remnant . The background regions were selected in a way to minimise the contamination caused by the 
PSF wings as well as from the stray light.  
The excess in the south of the FPMA image is the stray 
light from  X-rays that hit the detector without impinging on the optics \citep{wik14}.    We use the same source regions for Chandra observations.
The results of our study  of the  the spatial distribution of X-rays appeared quite  
similar the one reported by 
\citep{zoglauer15}.  Therefore, in this paper we do not discuss 
the morphology of the source  but focus on the 
study of spectral features of radiation.  

The spectral shape of synchrotron radiation in the cutoff region is sensitive to the spectrum of highest energy electrons which, in its turn,   depends on the electron acceleration and energy loss rates.  To  explore a  broad class of spectra, we describe  the spectrum of X-rays in the following  general form: 
\begin{equation}
\frac{{\rm d}N}{{\rm d} \epsilon} =
A E^{-\Gamma} \exp[-(\epsilon/\epsilon_0)^\beta_{\rm e}] \ .  
\label{spectrum}
\end{equation}
The change of the index $\beta$ 
in the second (exponential)  term allows 
a broad range of  spectral behaviour in  the cutoff region. For
example, $\beta=0$ implies a pure power-law distribution, while 
$\beta=1$ corresponds to  a power-law with a simple exponential cutoff. 

In the fitting procedure, in addition to the three parameters $\epsilon_0, \Gamma$ and $\beta$, one
should introduce  one more parameter,  the column density  $N_{\rm H}$,  
which takes into account the energy-dependent absorption of X-rays. 
We fix this parameter to the value  found by \citet{zoglauer15} from the fit of data by their {\it srcut} spectral model . Strictly speaking,  the best fit value of the column density should be different for different spectral model. To check the impact of different spectral models  on the  column density,  we adopted  different
functions  leaving the  column density  as a free parameter in the fitting procedure.  
We  found that the difference of the best fit column density  and the above fiducial value 
is less than several percent. Therefore, in order to keep the procedure  simple and minimise the number of free parameters,  we adopt the value  $N_{\rm H}=7.23 \times 10^{22}$   from the paper  of \citet{zoglauer15}. 
 
The results of our fit of the NuSTAR and Chandra spectral points 
using the model  ``power-law with exponential cutoff" in the general form of Eq.(1), i.e. 
leaving  $\beta$, $\Gamma$ and  $\epsilon$ as  free parameters,
are shown in  Table 1. One can see that the best fit gives a rather narrow range of the index $\beta$ around 1/2.  In Table we show separately also the results of the fits with three fixed values 
of $\beta$: 0, 1/2, and 1.  While the pure-power law  spectrum ($\beta=0$) can be unambiguously excluded, the model 
of power-law with a simple exponential cutoff ($\beta=1$)  is not favourable either. It is 
excluded  at the $3 \sigma$ statistical significance level.  In summary, the  combined Chandra and NuSTAR data are best described by the index $\beta_{\rm  e} \approx  0.5$ and $\epsilon_0 \approx 1.5$~keV.   

\begin{figure*}
\centering
\includegraphics[width=0.55\linewidth]{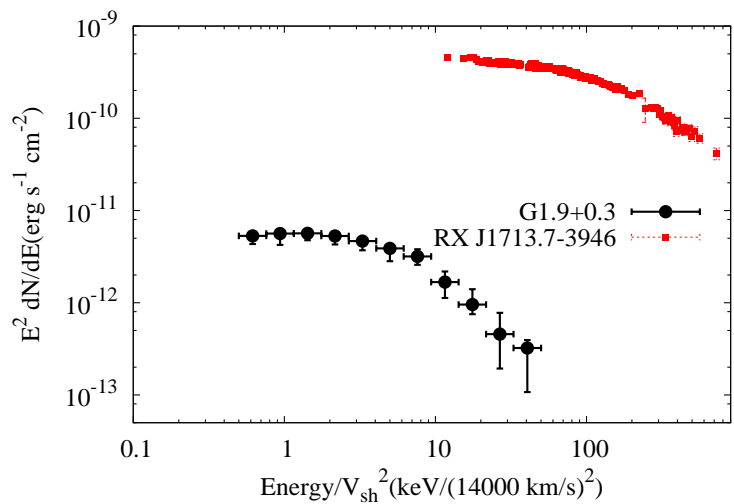}
\caption{The spectral points of G1.9+0.3 (this work; black circles)  
and  RX~J1713.4-3946 (red square) from \citet{1713_suzaku}. The energies 
of  the points of RX~J1713.4-3946 are rescaled 
by the  factor of the square of the ratio of shock speeds of J1713.4-3946  and  G1.9+0.3:
$\rm (14,000 \ km/s  / 4000 \ km/s)^2=12.25$.
}
\label{fig:sed1}
\end{figure*}

Whereas $\beta=1/2$ seems to be a natural  outcome (see below),  the cutoff energy around 1.5 keV   is  a rather unexpected result. 
Namely, it implies that the acceleration of electrons in G1.9+0.3 proceeds significantly slower than one would anticipate given the 
very large, 14,000 km/s  shock speed.  This can be seen from the comparison of 
of the SED of G1.9+0.3  with one of the most effective particle accelerators in our 
Galaxy,  $\approx$1600 year old SNR RX~J1713.4-3946 (see  Fig.\ref{fig:sed1}).
The cutoff energy in the synchrotron spectrum of shock-accelerated electrons 
is proportional to the square of  shock speed $v_{\rm sh}^2$ \citep{AhAt99}. Therefore,  in order to exclude the difference in the cutoff energies caused by  the difference in the shock speeds,  
we rescale the energies of the spectral points of  RX~J1713.4-3946  by  the factor $(v_{\rm sh}/14,000 \ \rm km/s)^2$, where the shock speed of  RX~J1713.4-3946 is about 
$v_{\rm sh} \simeq 4,000 \ \rm km/s$ \citep{Uchiyama07}. 
After 
such normalisation,  the cutoff energy of  
RX J1713.4-3946  becomes an order of magnitude higher than the cutoff in G1.9+0.3.
The acceleration of  electrons in RX J1713.4-3946 proceeds close to the 
Bohm diffusion limit thus provides an acceleration rate close to the maximum value\citep{Uchiyama07,zirakashvili10}. 
Consequently, we may conclude that the current acceleration rate of electrons in 
G1.9+0.3 is lower, by an order of magnitude,  compared to the maximum possible rate. 

It should be noted that  the physical meaning of 
Eq.(\ref{spectrum}) should not be overestimated. Namely, it 
should be considered as a
convenient analytical presentation of  the given set of 
measured spectral points. Consequently, 
the $\Gamma,\beta,\epsilon_0$) that enter into Eq.(\ref{spectrum}),  
should be treated as a combination of formal fit parameters 
rather than  physical quantities. 
For example,  $\epsilon_0$ in the exponential term of 
Eq.(\ref{spectrum})  should not  necessarily coincide with the cutoff energy (or 
maximum in the SED). Indeed, in different ($\Gamma,\beta,\epsilon_0$) combinations describing the same spectral points,
the parameter $\epsilon_0$ could have significantly different values. 
Analogously, $\Gamma$ should not be treated as a power-law index but rather 
a parameter which, in combination with $\Gamma$ and $\beta$, determines the slope 
(the tangential) of the spectrum  immediately before the cutoff region.  

\begin{figure*}
\centering
\includegraphics[width=0.55\linewidth]{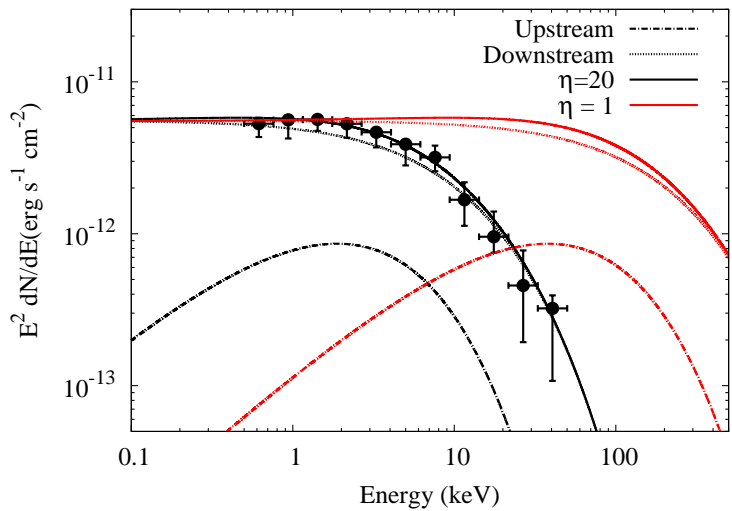}
\caption{The spectral points of G1.9+0.3 (this work) compared  to the predictions of synchrotron radiation of the shock accelerated electrons in the downstream and upstream regions \citep{zirakashvili07} for two regimes of diffusion: Bohm diffusion $\eta=1$ and 20 times faster, $\eta=20$. 
}
\label{fig:sed2}
\end{figure*}

The maximum acceleration rate of particles is 
achieved when it proceeds in the Bohm diffusion limit.  In the energy-loss dominated regime, the spectra of synchrotron radiation  can be expressed  by  
simple analytical  formulae  \citep{zirakashvili07}.  Because of compression of the magnetic field,
the overall synchrotron flux of the remnant is dominated by the 
radiation from the downstream region (see Fig.\ref{fig:sed2}). The SED of the latter   can be presented in the following form  \citep{zirakashvili07}
\begin{equation}
\epsilon^2\frac{{\rm d}N}{{\rm d} \epsilon} \propto 
\epsilon^2 (\epsilon / \epsilon_0)^{-1} [1+0.38  (\epsilon/\epsilon_0)^{0.5}]^{11/4} 
\exp[-(\epsilon/\epsilon_0)^{1/2}  \ .  
\label{shape}
\end{equation}
with   
\begin{equation}
\epsilon _0= \hbar \omega _0= \frac {\mathrm{2.2\ keV}}{\eta
(1+\kappa ^{1/2})^2}\left( \frac {u_1}{\mathrm{3000\ km\ }s^{-1}}
\right) ^2 \ , 
\label{e0}
\end{equation}
where  $\eta$ takes into account  
the deviation of the diffusion coefficient from its minimum value 
(in the nominal Bohm diffusion limit $\eta =1$). 
In the standard 
shock acceleration theory,   the momentum index of accelerated electrons 
$\gamma_{\rm s}=4$, and the ratio of the upstream and downstream 
magnetic fields, $\kappa=1/\sqrt{11}$.

In Fig.\ref{fig:sed2}  the  spectral points of G1.9+0.3
are compared with the theoretical predictions for synchrotron radiation in the 
upstream and downstream regions \citep{zirakashvili07}. The calculations are performed for 
two values of the parameter $\eta$  characterising the acceleration efficiency: $\eta=1$ 
(Bohm diffusion regime) and 20 times slower ($\eta=20$). The good (better than 20 \%) agreement of the spectral points with the theoretical curves 
for $\eta=20$ tells us that in G1.9+0.3 electrons are accelerated 
only at the 5 \% efficiency level. 

Although in the paper  of 
\citet{zoglauer15} the spectral points are not  explicitly presented, thus the direct 
comparison with our results is not possible,  the conclusions of our study on the energy spectrum of  G1.9+0.3 seems to be in agreement with the results of 
\citet{zoglauer15}. However, because of the  incorrect interpretation of the process of 
formation of the spectrum of synchrotron radiation, the statements in the paper by 
\citet{zoglauer15}  are misleading (see Appendix \ref{app:a}).  

 \begin{table*}[htbp]
\caption{Spectral Fitting results for G1.9+0.3 } \label{tab:1} \centering
\begin{tabular}{lllllll}
\hline
model&\vline ~PL index&\vline ~cutoff (keV)&\vline ~$\beta$ &\vline $\chi ^2 /d.o.f.$&\\
\hline
\hline
PL   &\vline ~2.54 (2.52 - 2.56) &\vline&\vline &\vline ~1089.4/666 &\\
\hline
PL+ecut   &\vline ~2.04 (1.98 - 2.10) &\vline~11.8 (10.5 - 13.3)&\vline &\vline ~697.7/665 &\\
\hline
PL+ecut ($\beta$=0.5)  &\vline ~1.65 (1.60 - 1.70)&\vline ~1.68 ( 1.50 - 1.90)&\vline ~0.5&\vline ~686.2/665\\
\hline
PL+ecut  ($\beta$ free)  &\vline ~1.62(1.48 - 1.75)&\vline ~ 1.41 (1.30-1.55)&\vline ~0.48 (0.40-0.56)&\vline ~685.8/664 &\\
\hline

\end{tabular}
\end{table*}

\section{Relativistic electrons and magnetic fields}

\begin{figure*}
\centering
\includegraphics[width=0.4\linewidth, height=0.3\linewidth]{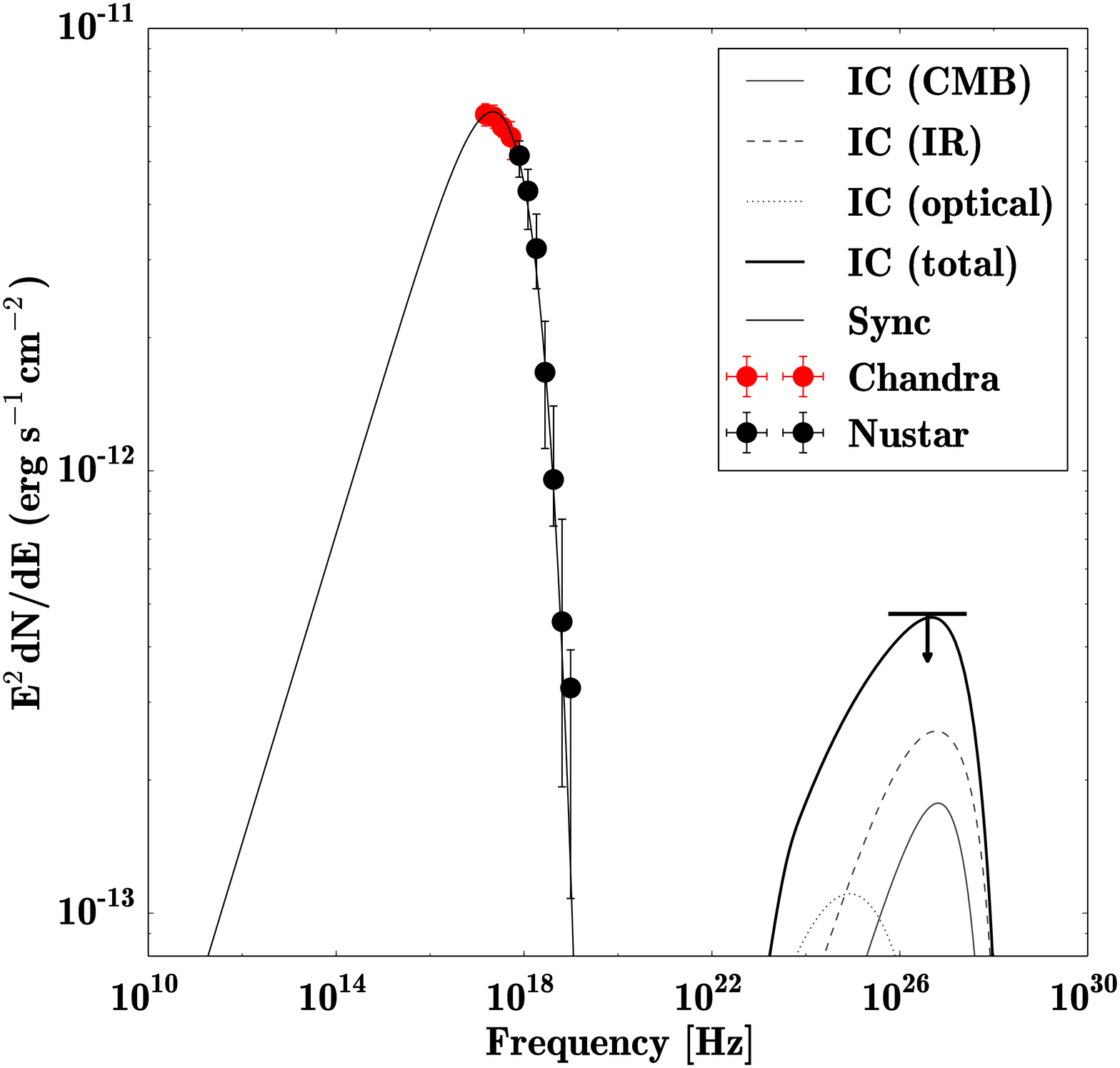}
\caption{X-ray SED as well as the VHE upper limit from \citet{hessG1.9}. The curves are the synchrotron and IC emissions fitted to derive the lower limit of the magnetic field.
}
\label{fig:SEDmodeling}
\end{figure*}

The joint treatment  of X-ray and $\gamma$-ray data, under the 
simplified assumption that the same electron population is responsible for 
the broad-band radiation through the synchrotron and inverse Compton channels, 
provides  information about the magnetic field and the total energy budget in relativistic electrons.   G1.9+0.3  has been observed in VHE 
$\gamma$-ray band with the  H.E.S.S.  Cherenkov telescope system. Although
no positive signal has been detected \citep{hessG1.9},   the $\gamma$-ray flux 
 upper limits allow meaningful constraints    on the 
the average magnetic field in the X-ray and $\gamma$-ray production region. For calculations of the broad-band SED, we adopt the same background radiation fields used in the paper \citet{hessG1.9}: the  infrared component with temperature of $48~\rm K$ and energy density of $1.5~\rm eV cm^{-3}$,  and the  optical component with temperature of $4300~\rm K$ and the energy density of $14.6~\rm eV cm^{-3}$. The comparison of   model calculations with observations (see Fig.\ref{fig:SEDmodeling}) give a  
lower limit of the magnetic field, $B \geq 17 \rm \mu G$.  

Under certain assumptions, the magnetic field can be constrained also based only 
on the X-ray data. In the ``standard" shock acceleration scenario, electrons are accelerated with  the  power-law index $\alpha =2$. However 
because of the short  radiative  cooling time, their spectrum of highest energy 
electrons (the X-ray producers)  becomes steeper, $\alpha=2 \to 3$. Consequently,  in the downstream region,  where the bulk of synchrotron radiation  is produced,  
X-rays have a photon index  $\Gamma=2$.    The synchrotron cooling time can be expressed through the magnetic field and the X-ray photon energy: 
$t_{\rm synch}  \simeq 50  (B/100 \rm \mu G)^{-3/2} (\epsilon/1~\rm keV)^{-1/2}$~years. Thus 
for $\epsilon \sim 1 \rm \ keV$ and the age of the SNR $\sim 150$~yr, we find that the magnetic field should be larger than $50 \mu$G. 
 
\begin{figure*}
\centering
\includegraphics[width=0.4\linewidth, height=0.3\linewidth]{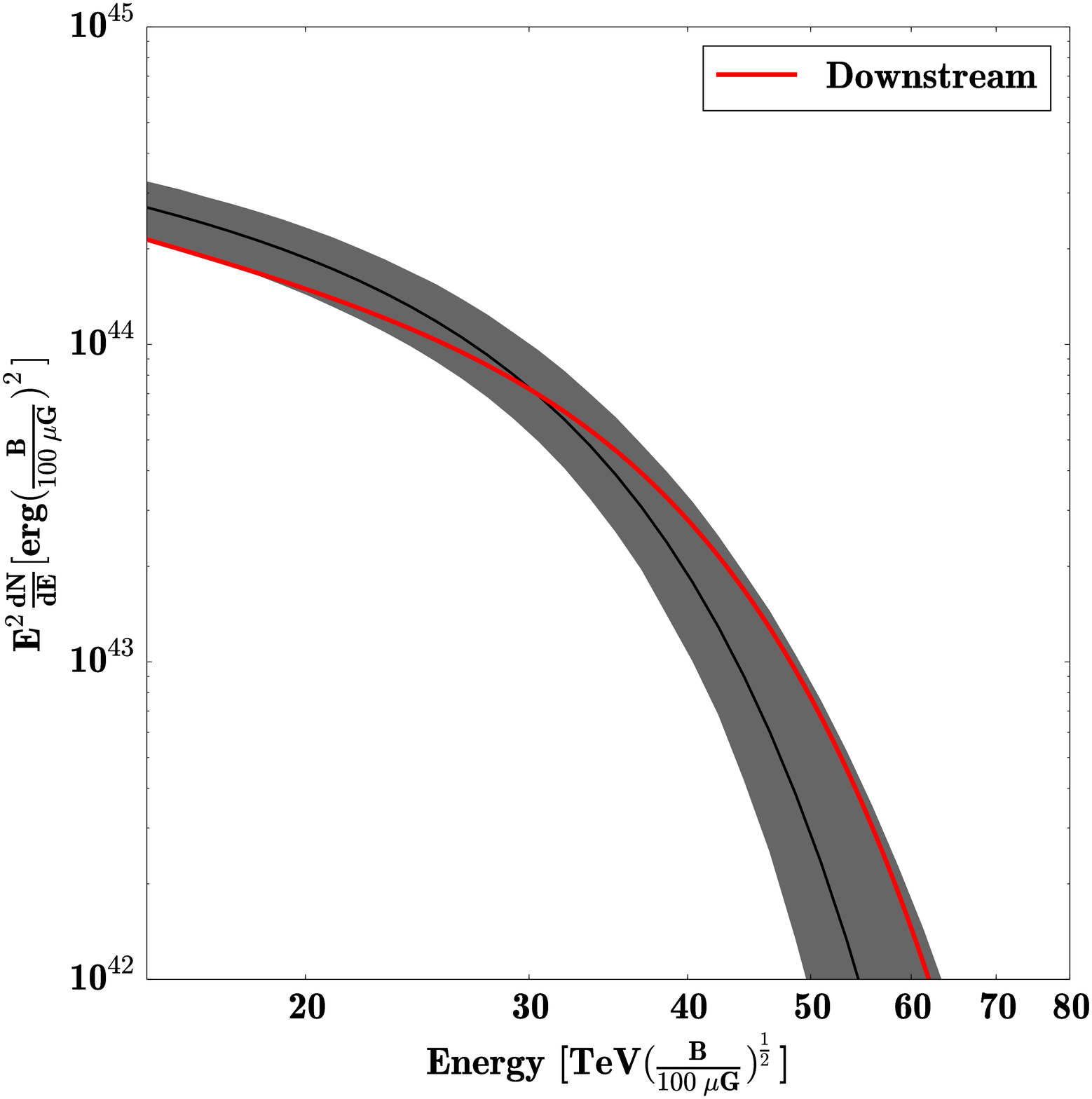}
\caption{Electron spectrum from the x-ray data points (black curve and  shaded area) and theoretical predicted integrated electron spectrum in young SNR (red curve) assuming a fast diffusion, i.e., $\eta = 20$ in Eq.1. Also shown is the contribution from  downstream region. %
}
\label{fig:ele}
\end{figure*}

The  combined Chandra and NuSTAR data  cover two decades in energy, 
from sub-keV to tens of keV.   This allows  derivation of the energy distribution 
of electrons, $W(E)=E^2{\rm d}N_{\rm e}/{\rm d}E$ in the most interesting region around the cutoff. 
The results shown in Fig.\ref{fig:ele}  are obtained using the  
Markov Chain Monte Carlo (MCMC) code {\it Naima} developed by V. Zabalza  
\footnote{\url{https://github.com/zblz/naima}}. It is assumed that the magnetic field is homogeneous both in space and time.  The results shown in
in Fig.\ref{fig:ele} are calculated for the  fiducial  value of  the magnetic field $B=100~\rm \mu G$, however they can be rescaled for any other value of the field. Note that while 
the shape of the spectrum  does not depend on the strength of the magnetic field, the energies of individual electrons scale as $E \propto B^{-1/2}$, and the total energy contained in electrons scales as $\propto B^{-2}$.  
Since in the ``standard"  
diffusive shock acceleration scenario the  synchrotron X-ray flux is contributed mainly by the downstream region,  the results in  Fig.\ref{fig:ele} correspond to the range of the 
energy distribution of electrons for the same region.  For comparison we show 
the energy distribution of  electron calculated 
using the formalism of \cite{zirakashvili07}. 
Apparently, the good agreement between the derived electron spectrum
with the theoretical curve  for $\eta$
naturally reflects the agreement between the X-ray observations and the theoretical predictions as demonstrated in Fig.\ref{fig:sed2}.

\section{Conclusions}
SNRs are believed to be the major contributors  to the Galactic CRs. The recent 
detections of TeV emission from more than ten young SNRs (of the age of  a few thousand 
years or younger),  demonstrates the ability of 
these objects  to  
accelerate  particles, electrons and/or protons, to energies up to 100 TeV. 
Yet,  we do not have observational evidence of extension of hard 
$\gamma$-ray  spectra  well beyond 10~TeV. Therefore one cannot claim 
an acceleration of protons and nuclei  by SNRs  to PeV energies.  On the other hand, 
one cannot claim the opposite either,  given the possibility that the acceleration of PeV protons and nuclei could happen at the early stages of evolution of SNRs when the shock speeds exceed  10,000~km/s. 
Then, the escape of the highest energy  particles 
at later stages  of evolution of SNRs can explain the spectral steepening of gamma-rays at multi-TeV energies  from  $\geq 1$ thousand years old  remnants.  

In this regard, the youngest known SNR in our Galaxy,  G1.9+0.3, with the measured shock speed 14,000~km/s, seems to be a unique object in our Galaxy to explorer the 
potential of SNRs for acceleration of protons and nuclei to PeV energies. Such measurements have been performed with the H.E.S.S. array of Cherenkov telescopes.  Unfortunately, no positive signal has been detected. On the other hand,
the recent observations of G1.9+0.3  in hard X-rays by NuSTAR provide unique information about the acceleration efficiency of electrons. Together with Chandra data at lower energies,  these data allow model-independent conclusions.  Although the general shape of the energy spectrum of X-rays is  in a very good agreement with predications of the diffusive shock-acceleration theory, the acceleration rate appears an order of magnitude slower relative to the maximum acceleration rate achieved in the nominal Bohm diffusion limit. To a certain extent, this is a surprise, especially when compared with young SNRs like Cas A and 
RX~J1713.4-3946 in which the acceleration of electrons proceeds in the regime close to the Bohm diffusion.  If the acceleration of protons and nuclei proceeds in the same manner as the electron acceleration, this result  could have  a negative  impact on the ability of G1.9+0.3 to operate as a PeVatron. Apparently, the observations of G1.9+0.3 alone are not sufficient to conclude whether this  conclusion can be generalised for other SNRs.

\appendix

\section{}
\label{app:a}
The cutoff region in the distribution of parent electrons $F(E)$  has a shape similar to the synchrotron spectrum given by Eq.(\ref{shape}),  
$F(E) \propto \exp[-(E/E_0)^\beta_{\rm e}]$   with the following simple relation between $\beta_{\rm e}$ and $\beta$ \citep{Fritz89}:
\begin{equation}
\beta=\frac{\beta_{\rm e}}{2+\beta_{\rm e}} \ .
\end{equation}
Thus the cutoff region in the spectrum of synchrotron radiation 
is much smoother  than the cutoff region in of the spectrum of parent electrons. 
For any electron distribution, the the synchrotron cutoff cannot be sharper than 
the simple exponential decline ($\beta=1$) which can be  realised only in the case of 
an abrupt cutoff in the spectrum of parent electrons  ($\beta_{\rm e} \to \infty$).   
In the case of a simple exponential cutoff in the electron spectrum ($\beta_{\rm e}=1$), 
the corresponding synchrotron cutoff region is very shallow with $\beta=1/3$. Formally, such spectra can be formed during the shock acceleration of electrons in the Bohm diffusion regime when the maximum energy of electrons is determined by the age of the source rather than by energy losses of electrons. However,  
in the particular case of 
young SNRs when the electrons  are accelerated up to 100 TeV and  beyond (otherwise one cannot explain  the observed X-ray  data), any reasonable set of parameters (magnetic field, age of the source, shock speed, {\it etc.}), the acceleration proceeds in the electron energy-loss regime, and the maximum energy is determined from the competition between the acceleration and energy loss rates.  In this case,
the spectrum of electrons  exhibits  a super exponential cutoff, namely $\beta_{\rm e}=2$  \citep{zirakashvili07}. Correspondingly, the synchrotron spectrum contains, in accordance with Eq.(\ref{e0}),  a cutoff with $\beta=1/2$.  

The  
$\delta$-functional approximation gives wrong relation between $\beta_{\rm e}$ and $\beta$, namely  $\beta = \beta_{\rm e}/2$. For example, within this approximation, the 
cutoff in the spectrum of synchrotron radiation with $\beta=1$ can be interpreted as a
result of the cutoff in the electron spectrum with $\beta_{\rm e}=2$.
However, in reality,  such a synchrotron spectrum can be formed, 
in accordance with Eq.(\ref{shape}), in the case of an abrupt cutoff in the electron spectrum, 
i.e.  $\beta_{\rm e} \to \infty$.
 Another remark  regarding the {\rm srcut} model. 
 The cutoff in the synchrotron spectrum with $\beta=1/2$ is formed by 
the highest energy electrons with $\beta_{\rm e}=2$  (see Eq.(2)) but not with 
 $\beta_{\rm e}=1$, as it follows from the $\delta$-functional approximation. 
Therefore the often claims in the literature that the $\delta$-functional 
approximation gives  correct synchrotron spectrum  \citep[see, e.g., ][]{Reynolds2008}
 is not only wrong, but  leads to  misleading conclusions.  
 For example,  while the fit of the X-ray data by a {\it script} type spectrum is a clear indication of  acceleration of electrons in the energy-loss regime, the same result has been  interpreted by \citep{zoglauer15}  as acceleration in the regime free of energy losses.
 
\bibliographystyle{aa}
\bibliography{ms}

\begin{thebibliography}{19}
\expandafter\ifx\csname natexlab\endcsname\relax\def\natexlab#1{#1}\fi

\bibitem[{{Aharonian}(2013)}]{Aharonian2013}
{Aharonian}, F.~A. 2013, Astroparticle Physics, 43, 71

\bibitem[{{Aharonian} \& {Atoyan}(1999)}]{AhAt99}
{Aharonian}, F.~A. \& {Atoyan}, A.~M. 1999, \aap, 351, 330

\bibitem[{{Bamba} {et~al.}(2008){Bamba}, {Fukazawa}, {Hiraga}, {Hughes},
  {Katagiri}, {Kokubun}, {Koyama}, {Miyata}, {Mizuno}, {Mori}, {Nakajima},
  {Ozaki}, {Petre}, {Takahashi}, {Takahashi}, {Tanaka}, {Terada}, {Uchiyama},
  {Watanabe}, \& {Yamaguchi}}]{sn1006_suzaku}
{Bamba}, A., {Fukazawa}, Y., {Hiraga}, J.~S., {et~al.} 2008, \pasj, 60, S153

\bibitem[{{Blasi}(2013)}]{blasi13}
{Blasi}, P. 2013, \aapr, 21, 70

\bibitem[{{Fritz}(1989)}]{Fritz89}
{Fritz}, K.~D. 1989, \aap, 214, 14

\bibitem[{{Green} {et~al.}(2008){Green}, {Reynolds}, {Borkowski}, {Hwang},
  {Harrus}, \& {Petre}}]{green08}
{Green}, D.~A., {Reynolds}, S.~P., {Borkowski}, K.~J., {et~al.} 2008, \mnras,
  387, L54

\bibitem[{{H.E.S.S.~Collaboration} {et~al.}(2014){H.E.S.S.~Collaboration},
  {Abramowski}, {Aharonian}, {Benkhali}, {Akhperjanian}, {Ang{\"u}ner},
  {Anton}, {Balenderan}, {Balzer}, {Barnacka}, \& et~al.}]{hessG1.9}
{H.E.S.S.~Collaboration}, {Abramowski}, A., {Aharonian}, F., {et~al.} 2014,
  \mnras, 441, 790

\bibitem[{{Hillas}(2013)}]{Hillas2013}
{Hillas}, A.~M. 2013, Astroparticle Physics, 43, 19

\bibitem[{{Malkov} \& {Drury}(2001)}]{Misha}
{Malkov}, M.~A. \& {Drury}, L.~O. 2001, Reports on Progress in Physics, 64, 429

\bibitem[{{Reynolds}(2008)}]{Reynolds2008}
{Reynolds}, S.~P. 2008, \araa, 46, 89

\bibitem[{{Reynolds} {et~al.}(2008){Reynolds}, {Borkowski}, {Green}, {Hwang},
  {Harrus}, \& {Petre}}]{reynolds08}
{Reynolds}, S.~P., {Borkowski}, K.~J., {Green}, D.~A., {et~al.} 2008, \apjl,
  680, L41

\bibitem[{{Reynolds} {et~al.}(2009){Reynolds}, {Borkowski}, {Green}, {Hwang},
  {Harrus}, \& {Petre}}]{reynolds09}
{Reynolds}, S.~P., {Borkowski}, K.~J., {Green}, D.~A., {et~al.} 2009, \apjl,
  695, L149

\bibitem[{{Stage} {et~al.}(2006){Stage}, {Allen}, {Houck}, \&
  {Davis}}]{stage06}
{Stage}, M.~D., {Allen}, G.~E., {Houck}, J.~C., \& {Davis}, J.~E. 2006, Nature
  Physics, 2, 614

\bibitem[{{Tanaka} {et~al.}(2008){Tanaka}, {Uchiyama}, {Aharonian},
  {Takahashi}, {Bamba}, {Hiraga}, {Kataoka}, {Kishishita}, {Kokubun}, {Mori},
  {Nakazawa}, {Petre}, {Tajima}, \& {Watanabe}}]{1713_suzaku}
{Tanaka}, T., {Uchiyama}, Y., {Aharonian}, F.~A., {et~al.} 2008, \apj, 685, 988

\bibitem[{{Uchiyama} {et~al.}(2007){Uchiyama}, {Aharonian}, {Tanaka},
  {Takahashi}, \& {Maeda}}]{Uchiyama07}
{Uchiyama}, Y., {Aharonian}, F.~A., {Tanaka}, T., {Takahashi}, T., \& {Maeda},
  Y. 2007, \nat, 449, 576

\bibitem[{{Wik} {et~al.}(2014){Wik}, {Hornstrup}, {Molendi}, {Madejski},
  {Harrison}, {Zoglauer}, {Grefenstette}, {Gastaldello}, {Madsen},
  {Westergaard}, {Ferreira}, {Kitaguchi}, {Pedersen}, {Boggs}, {Christensen},
  {Craig}, {Hailey}, {Stern}, \& {Zhang}}]{wik14}
{Wik}, D.~R., {Hornstrup}, A., {Molendi}, S., {et~al.} 2014, \apj, 792, 48

\bibitem[{{Zirakashvili} \& {Aharonian}(2007)}]{zirakashvili07}
{Zirakashvili}, V.~N. \& {Aharonian}, F. 2007, \aap, 465, 695

\bibitem[{{Zirakashvili} \& {Aharonian}(2010)}]{zirakashvili10}
{Zirakashvili}, V.~N. \& {Aharonian}, F.~A. 2010, \apj, 708, 965

\bibitem[{{Zoglauer} {et~al.}(2015){Zoglauer}, {Reynolds}, {An}, {Boggs},
  {Christensen}, {Craig}, {Fryer}, {Grefenstette}, {Harrison}, {Hailey},
  {Krivonos}, {Madsen}, {Miyasaka}, {Stern}, \& {Zhang}}]{zoglauer15}
{Zoglauer}, A., {Reynolds}, S.~P., {An}, H., {et~al.} 2015, \apj, 798, 98

\end{thebibliography}
\end{document}